\documentstyle[prb,aps,psfig]{revtex}
\begin{document}
\draft
\twocolumn[\hsize\textwidth\columnwidth\hsize\csname 
@twocolumnfalse\endcsname 

\title{Electronic transport and the localization length in the quantum 
Hall effect}
\author{M. Furlan}
\address{Swiss Federal Institute of Technology EPFL, CH-1015 Lausanne, 
Switzerland; and\\
Swiss Federal Office of Metrology OFMET, CH-3084 Wabern, Switzerland}
\date{\today}
\maketitle 

\begin{abstract}
We report on recent experimental results from transport measurements with 
large Hall bars made of high mobility GaAs/AlGaAs heterostructures.  Thermally 
activated conductivities and hopping transport were investigated in the integer 
quantum Hall regime.  The predominant transport processes in two dimensions are 
discussed.  The implications of transport regime on prefactor universality and on 
the relation between $\rho_{\mathrm{xx}}$ and $\rho_{\mathrm{xy}}$ are studied.  
Particularly in the Landau level tails, strictly linear dependence 
$\delta\rho_{\mathrm{xy}}(\rho_{\mathrm{xx}})$ was found, with pronounced 
asymmetries with respect to the plateau centre.  At low temperatures, Ohmic 
(temperature dependent) as well as non-Ohmic (current dependent) transport were 
investigated and analysed on the basis of variable-range hopping theory.  The 
non-Ohmic regime could successfully be described by an effective electron 
temperature model.  The results from either the Ohmic transport or from a 
comparison of Ohmic and non-Ohmic data allowed to determine the localization 
length $\xi$ in two different ways.  The observed divergence of $\xi(\nu)$ with 
the filling factor $\nu$ approaching a Landau level centre, is in qualitative 
agreement with scaling theories of electron localization.  The absolute values of 
$\xi$ far from the $\rho_{\mathrm{xx}}$ peaks are compared with theoretical 
predictions.  On one hand, discrepancies between the $\xi$ results obtained from 
the two experimental methods are attributed to an inhomogeneous electric field 
distribution.  Extrapolation yields an effective width of dominant potential drop 
of about $100\ \mu$m.  On the other hand, our analysis suggests a divergence of 
the dielectric function $\epsilon_{\mathrm{r}} \propto \xi^{\beta}$ with 
$\beta \simeq 1$.
\end{abstract}

\pacs{PACS numbers: 73.40.Hm, 73.20.F}
\vskip5pc 
] 
%

\section{Introduction}
\label{foreplay}
Transport measurements in the quantum Hall\cite{QHE} regime have been 
widely used to investigate fundamental physics of electron conduction 
in the case of a quantized two-dimensional electron gas in strong 
perpendicular magnetic fields.  If the energy separation between the 
Landau levels (LL), i.e.  the cyclotron energy $\hbar 
\omega_{\mathrm{c}}$, is much larger than the LL linewidth, all 
electrons within the LL tails are considered to be localized.  The 
localization length $\xi$ characterizes the size of the space region 
in which the wave function of an electron, moving in an impurity 
potential, is not exponentially small.  This characteristic length is 
believed to diverge with the Fermi level $E_{\mathrm{F}}$ approaching 
the LL centre.  In that limit, the divergence can be expressed 
according to a power law:
\begin{equation}
  \xi(\nu) \propto |\nu - \nu_{\mathrm{c}}|^{-\gamma},
  \label{xidiverg}
\end{equation}
where $\nu = 2 \pi \ell_{\mathrm{B}}^2 n_{\mathrm{e}}$ is the filling 
factor (with $\ell_{\mathrm{B}} = \sqrt{\hbar / e B}$ the magnetic 
length and $n_{\mathrm{e}}$ the electron sheet density).  The value 
$\nu_{\mathrm{c}}$ corresponds to the position where $E_{\mathrm{F}}$ 
coincides with the LL centre, and $\gamma \simeq 2.3$ is a critical 
exponent.\cite{Wei4,Pruisken,critExp,KochScal} On the 
other hand, the localization length at the resistivity minima was 
recently predicted\cite{loclength} to be on the order of the classical 
cyclotron radius (true for our samples, although depending on the phase 
diagram introduced in Ref.~\onlinecite{loclength}).  The transport 
properties depend on disorder and on the temperature.  Different 
dominant transport processes are distinguished for different 
temperature ranges.  At intermediate temperatures (typically a few 
Kelvin), conductance is predominantly determined by electrons 
thermally activated to the nearest extended states.  The diagonal 
conductivity tensor component $\sigma_{\mathrm{xx}}$ then follows an 
Arrhenius Law
\begin{equation}
  \sigma_{\mathrm{xx}}(T) = \sigma_{\mathrm{xx}}^0\, 
  {\mathrm{e}}^{\, - T_{0} / T}.
  \label{EqTactiv}
\end{equation}
The activation energy $k_{\mathrm{B}} T_{0}$ corresponds to the 
distance between the Fermi energy $E_{\mathrm{F}}$ and the percolation 
level $E_{\mathrm{c}}$.  In several 
works,\cite{moreprefac,PS-short,PS-pre,FPS} the universality and 
possible dependencies of the prefactor values $\sigma_{\mathrm{xx}}^0$ 
were studied.

Based on considerations of localization and percolation (and in the limit 
$\sigma_{\mathrm{xx}} \ll e^2/h$), a similar thermally activated 
behaviour is expected for the deviation 
\[\delta\sigma_{\mathrm{xy}}(T) = \sigma_{\mathrm{xy}}(T) - 
\frac{e^2 \nu_{0}}{h}\] 
from the quantized Hall conductivity, where $\nu_{0}$ is 
the integer filling factor at the plateau centre.  In spite of the 
importance to understand the reasons and processes leading to a 
possible deviation from the quantized values, a clear and unambiguous 
distinction between pure thermal activation and other effects (e.g.  
sample dependent mixing of the tensor components) has not yet been 
experimentally achieved.  At integer filling factors and sufficiently 
wide cyclotron gaps, however, a linear dependence 
$\delta\sigma_{\mathrm{xy}}(T) \propto \sigma_{\mathrm{xx}}(T)$ has 
often been observed.\cite{LinDep} Unfortunately, such 
experiments have not been extended to the Hall plateau regions further 
away from the plateau centres.  In the dissipative regime of plateau 
transitions, where the Shubnikov-de Haas (SdH) peaks emerge, the 
situation is again different: $\sigma_{\mathrm{xy}}$ and 
$\sigma_{\mathrm{xx}}$ are not independent variables, but are 
described by a two-parameter renormalization-group 
theory,\cite{Pruisken,Wei3} satisfying the so-called ``semicircle 
rule".\cite{semicircle}

Whereas a lot of experiments were concerned with measurements at the 
$\sigma_{\mathrm{xx}}$ minima only, 
others\cite{Wei4,KochScal,Wei3,J-scal} specially 
concentrated on the transition regions.  In spite of large theoretical 
efforts, experiments to link the different regimes in order to 
complete the picture of thermally activated transport in the quantum 
Hall effect are still missing.

With decreasing temperature, the longitudinal conductivity becomes 
exponentially small, and an electron conduction mechanism known as 
variable-range hopping (VRH) becomes the dominant transport 
process.\cite{ElProp}  
Several attempts to describe the measured $\sigma_{\mathrm{xx}}(T)$ 
behaviour in the hopping regime with
\begin{equation}
  \sigma_{\mathrm{xx}} \propto {\mathrm{e}}^{-(T_{1} / T)^\alpha}
  \label{Eqgenhop}
\end{equation}
were reported.\cite{Eberthopping,Briggs} While Eq.~(\ref{Eqgenhop}) 
describes the Mott hopping\cite{Mott} with $\alpha=\frac{1}{3}$ 
$(\frac{1}{4})$ in two (three) dimensions in the absence of a Coulomb 
gap, a suppression of the density of states near the Fermi level due 
to Coulomb interactions\cite{Pollak} leads to a soft Coulomb 
gap,\cite{Efros} and expression (\ref{Eqgenhop}) with an exponent 
$\alpha=\frac{1}{2}$ was derived:
\begin{equation}
  \sigma_{\mathrm{xx}}(T) = \sigma_{\mathrm{xx}}^{\mathrm{T}}\, 
  {\mathrm{e}}^{-\sqrt{T_{1} / T}},
  \label{EqVRH}
\end{equation}
where
\begin{equation}
  k_{\mathrm{B}} T_{1}(\nu) = C\, \frac{e^2}{4\pi \epsilon_{\mathrm{r}} 
  \epsilon_{0} \xi(\nu)},
  \label{EqT1}
\end{equation}
with the numerical constant $C \approx 6.2$ in two 
dimensions,\cite{Nguyen} the dielectric function 
$\epsilon_{\mathrm{r}} \approx 13$ (value for GaAs), and the vacuum 
permittivity $\epsilon_{0}$.  The hopping behaviour (\ref{EqVRH}) was 
also derived in Refs.~\onlinecite{Ono,Wysokinski,Grunwald}, although with 
different coefficients for $T_{1}$ and 
$\sigma_{\mathrm{xx}}^{\mathrm{T}}$.  The role of the prefactor 
$\sigma_{\mathrm{xx}}^{\mathrm{T}}$, i.e.  whether it is temperature 
dependent or not, was widely disputed and still remains an unsolved 
problem.  Experimentally, a prefactor proportionality 
$\sigma_{\mathrm{xx}}^{\mathrm{T}} \propto 1/T$ was usually 
observed.\cite{Eberthopping,Briggs,KochHop,CoulGap}

Although the quantum Hall effect can successfully be described by 
means of the linear-response theory at low current levels, the 
non-Ohmic transport observed at high electric fields is not yet well 
understood.  Different models were proposed to explain the behaviour 
in the current region below the critical breakdown current: inter- and 
intra-LL 
transitions due to high local electric field\cite{interLL} (tunneling 
or emission of phonons), increase in the number of delocalized states 
in the LL\cite{trugman} or the production of superheated 
electrons.\cite{ehot} Experimentally, an 
essentially exponential dependence of the longitudinal resistivity on 
current has most often been observed.  In a recent model\cite{PS-peak} 
the non-Ohmic transport in the quantum Hall regime was discussed on the 
basis of the theory of hopping in a strong electric 
field.\cite{hopEfield} At low temperatures, non-Ohmic transport in the 
VRH regime is then expected to show a behaviour like
\begin{equation}
  \sigma_{\mathrm{xx}}(J) = \sigma_{\mathrm{xx}}^{\mathrm{J}}\, 
  {\mathrm{e}}^{-\sqrt{{\mathcal{E}}_{1} / {\mathcal{E}}_{\mathrm{H}}}},
  \label{EqJhop}
\end{equation}
where ${\mathcal{E}}_{\mathrm{H}}$ is the Hall electric field across 
the sample, and
\begin{equation}
  {\mathcal{E}}_{1} = \frac{2 k_{\mathrm{B}} T_{1}}{e\, \xi}
  \label{EHchar}
\end{equation}
is a characteristic value related to the hopping temperature.  This 
electric field dependent hopping transport model is based on the idea 
of the existence of a quasi-Fermi-level tilted by the electric field.  
As a consequence, the local Fermi distribution is formed corresponding 
to an effective temperature $T_{\mathrm{eff}} \propto 
{\mathcal{E}}_{\mathrm{H}} \xi$.  Hence, in analogy to the Ohmic 
hopping transport (\ref{EqVRH}) in the quantum Hall regime, the 
non-Ohmic conductivity in the limit of vanishing temperature is then 
immediately given by Eq.~(\ref{EqJhop}) for increasing electric field.

Both the temperature and current dependent VRH conductivities in 
Eqs.~(\ref{EqVRH}) and (\ref{EqJhop}) explicitly depend on the 
localization length $\xi$ via the characteristic values $T_{1}$ and 
${\mathcal{E}}_{1}$ (\ref{EqT1},\ref{EHchar}).  To test the 
predictions, it is interesting to compare experimental results (namely 
the extracted $\xi$) on the basis of the discussed hopping theory.  While 
Eq.~(\ref{EqT1}) also contains a dependence on 
$\epsilon_{\mathrm{r}}$, the localization length in (\ref{EHchar}) is 
a pure function of characteristic values obtained from experiment: a 
comparison of the measured conductivities $\sigma_{\mathrm{xx}}(T)$ 
and $\sigma_{\mathrm{xx}}(J)$ relates the current $J$ to an effective 
temperature like
\begin{equation}
  k_{\mathrm{B}} T_{\mathrm eff}(J) = e \xi\, 
\frac{\rho_{\mathrm{xy}} 
  J} {2 L_{\mathrm{y}}},
  \label{EqTeff}
\end{equation}
where $L_{\mathrm{y}}$ is the sample width.  The localization length 
$\xi$ can therefore be determined without explicitly knowing the exact 
behaviour of the prefactors.

In the present paper we report on experimental results from extensive 
transport measurements, covering the ranges of thermally activated and 
VRH transport in the integer quantum Hall effect.  The subsequent text 
is organized as follows: the experimental conditions and the sample 
properties are presented in Sec.~\ref{setup}.  The experimental 
results in Sec.~\ref{results} are divided in three parts.  First, the 
important results from thermally activated transport measurements are 
summarized, and the relations between the resistivity tensor components are 
discussed.  The other two parts are devoted to either temperature or 
current dependent measurements in the VRH regime.  Section 
\ref{discuss} gives a discussion about the determination of $\xi(\nu)$. 
Possible reasons for discrepancies between results, obtained from 
different experiments, are considered.  Conclusions are given in 
Sec.~\ref{conclude}.

\section{Sample properties and measurement technique}
\label{setup}
For the transport experiments, we have used Hall bars made of 
GaAs/AlGaAs heterostructures.  These are considered as ``good'' 
samples from the metrological point of view: they have wide Hall 
voltage plateaus ($\Delta B \approx 2$~T for plateau two), high critical 
breakdown currents ($J_{\mathrm{c}} > 10^{-4}$~A), and low contact 
resistances (typically well below 1~$\Omega$).  Nonideal contacts, 
i.e.  such with high resistance, introduce a nonequilibrium edge-bulk 
electron distribution.  At short distances and in the linear low 
current regime, such a nonequilibrium situation leads to the 
observation of the reported nonlocal 
resistances.\cite{selective,Crump} In this 
context, it is well established that high contact quality is crucial 
for the observability of the exact 
quantization\cite{HighPrec,Komiyama} as well as for correct 
determination of the longitudinal resistivity in Hall bars.  Our 
samples are also rather large (bar width and contact distances 
typically 1 mm) in order to omit narrow channel effects or poor 
equilibration between the probes.  We are stressing these facts to 
emphasize that our samples did not suffer nonlocal transport, in the 
sense as described in Ref.~\onlinecite{HaugEdge}.

For the present investigations, we concentrate on two samples 
from different production sources and with different properties, as 
listed in Table \ref{samples}.  Further details on the samples may also 
be found in Ref.~\onlinecite{HighPrec}.  Each Hall bar had three 
equidistant voltage probes on each side, which, upon injection of a dc 
current $J$, were all measured simultaneously, yielding the potential 
drop $V_{\mathrm{ij}}$ across any contact pair combination.  The 
resistances $R_{\mathrm{ij}}=V_{\mathrm{ij}}/J$ were determined from 
the mean value from the measurements with both current polarities.  
Thus, with the dc technique, thermoelectric effects or any instrumental 
offsets were safely canceled, while the information was still 
available to account for current path dependent effects: at very low 
current levels (typically below $10^{-8}$~A in our samples), nonlinear 
and in some cases strongly asymmetrical behaviour can be 
observed.\cite{lowJ} This low current regime was excluded from our data 
analysis, and we will only discuss transport either at current levels 
where linear response is applicable or in the non-Ohmic regime at 
higher currents.

The experiments were performed by varying current $J$, temperature $T$ 
and magnetic field density $B$.  We want to restrict the present 
discussion to the range of high magnetic field densities corresponding 
to filling factors within $1.5 < \nu < 4.5$, and particularly 
concentrate on the plateau regions.  According to this range, the 
plateaus around the filling factors $\nu \approx 2, 3, 4$ are called 
plateau two, three and four, respectively.  The spin-split LL with the 
index $N=1$, leading to the third plateau, was well resolved at low 
temperatures, i.e.  negligible overlap of the energy bands.  Two types 
of experiments were performed extensively: (a) at fixed $\nu$ the 
current was set to $|J| = 1\ \mu\mathrm{A}$ and the temperature varied 
in the range 300 mK $\leq T \leq$ 20 K, or (b) the temperature was 
kept constant at $T=324 \pm 12$ mK and the current varied in the range 
1 $\mu$A $\leq |J| \leq$ 100 $\mu$A.  The temperature measurement was 
performed with a Speer resistor and a capacitance thermometer, both 
calibrated on the basis of a germanium thermometer at $B=0$ and then 
extended with the necessary corrections due to magnetoresistive 
effects.  The accuracy of our thorough temperature measurement was 
comparable to the uncertainties given for the calibration curve (from 
Lakeshore) of the germanium sensor.  In experiment (a) the temperature 
was swept very slowly (typically several hours for a change in 
temperature by one order of magnitude) in order to guarantee thermal 
equilibration as well as sufficient statistics in the experimental 
data.  No hysteresis or other variations between consecutive $T$ 
sweeps could be observed.  In case (b) special care has been taken 
during the experiments as well as in the offline data analysis to keep 
the dissipation at a negligible level, i.e.  no significant 
temperature variation of the $^3$He bath.

The sample contact resistances were 
periodically controlled during the experiments in order to monitor any 
deviations of the sample quality that could lead to systematic errors.  
Upon proper and careful sample handling the experiments could be 
continuously carried out during several days with perfect 
reproducibility and with no significant change in sample properties.  
This was an essential condition to allow comparison of the data from 
different runs, since thermal cycling generally also changes sample 
characteristics, like the mean electron sheet density and the local 
charge and potential distribution.

From the measured longitudinal and transverse resistances 
$R_{\mathrm{L}}$ and $R_{\mathrm{H}}$, respectively, and with the usual 
assumption of a homogeneous sample, we determined the resistivity tensor 
components $\rho_{\mathrm{xx}} = \rho_{\mathrm{yy}} = R_{\mathrm{L}} 
L_{\mathrm{y}}/L_{\mathrm{x}}$ and $\rho_{\mathrm{xy}} = -\rho_{\mathrm 
yx} = R_{\mathrm{H}}$, with the distance $L_{\mathrm{x}}$ between the 
voltage probes and the width $L_{\mathrm{y}}$ of the Hall bar.  The 
conductivity tensor is given by $\sigma_{\mathrm{\mu\nu}} = 
(\rho^{-1})_{\mathrm{\mu\nu}}$.  We are presenting results for both 
quantities, depending on the theories referred to and the appropriate 
range of values considered.

\section{Experimental results}
\label{results}
In order to make sure that the temperature dependent experiments were 
not influenced by electron heating effects, we performed measurements 
of SdH oscillations at our lowest $^3$He bath temperature (300 mK) and 
varied the bias current.  Figure \ref{figSdH} shows such SdH traces 
$\rho_{\mathrm{xx}}(B)$ for sample A with current levels 30 nA $\leq J 
\leq$ 30 $\mu$A, together with the corresponding Hall resistivities 
$\rho_{\mathrm{xy}}(B)$.  A clear decrease of the SdH peaks is 
observed with decreasing $J$.  While the oscillations strongly depend 
on $J$ at high current levels, they saturate at about $J \lesssim 1\ 
\mu$A.  Therefore, we have chosen $J = 1\ \mu$A as a compromise 
between negligible electron heating and maximum sensitivity to low 
resistivities.

Also apparent in Fig.~\ref{figSdH} is the typical strong asymmetry 
between the SdH peaks corresponding to the up and down spin-split 
LLs.\cite{HaugAsym,Svoboda-coupl} This effect is qualitatively well 
explained by the 
theory\cite{Svoboda-coupl,Nachtwei-nonloc} of 
different equilibration probabilities of the edge and bulk channels at 
the sample edges due to different distances in the confining 
potential.  An increased ${\mathcal{E}}_{\mathrm{H}}$ tilts the 
potential and therefore reduces the channel distances.  However, the 
application of the theoretical 
formulation\cite{Svoboda-coupl,Nachtwei-nonloc} to 
our data yields equilibration lengths, which are by orders of 
magnitude larger than our sample size.  Furthermore, we see no 
difference in the resistivities measured at different contacts and at 
different distances.  The inadequacy of the theory to extract, in some 
cases, a physically meaningful equilibration length has already been 
pointed out in Ref.~\onlinecite{Crump}.

Another argument for the observed asymmetries of SdH peaks was given in 
Ref.~\onlinecite{HaugAsym}, according to which the DOS becomes 
asymmetric due to an unequal contribution of attractive and repulsive 
scatterers.  From our data analysis (cf.  note in Sec.~\ref{subVRH}), 
we could not find any significantly asymmetric DOS in our samples.  
Hence we have to reject in our case the picture of DOS correlated 
asymmetries of the measured SdH peaks.

\subsection{Thermal activation}
\label{subTactiv}
We have recently reported on experimental results of thermally 
activated longitudinal conductivities for our high mobility samples in 
the quantized plateau regimes.\cite{ep2ds} Clear activated behaviour, 
according to Eq.~(\ref{EqTactiv}), was observed in an intermediate 
temperature range $T \approx 1 \ldots 10$~K and typically over at 
least two decades in $\sigma_{\mathrm{xx}}$.  Measurements with lower 
bias currents by one order of magnitude led to the same results of the 
activated data, justifying the neglect of electron heating due to 
sufficiently low current density.  The activation energies 
$k_{\mathrm{B}} T_{0}$ were extracted from fitting 
Eq.~(\ref{EqTactiv}) to the maximum slopes of the data points.  At 
$\nu_{0}=2$ and $\nu_{0}=4$, they were found to be equal to half of 
the LL spacing $\hbar \omega_{\mathrm{c}} / 2$ within experimental 
uncertainties.  This is in perfect agreement with expectations and with 
the fact of negligible spin energy $g_{0}\mu_{\mathrm{B}}B \ll \hbar 
\omega_{c}$.  At $\nu_{0}=3$, increased activation energies compared 
to the bare spin-splitting energy were observed due to an effectively 
enhanced $g$-factor $g^{*} \approx 3.5 \ldots 5.4$ as a result of 
exchange interaction\cite{gfactor} (larger $g^{*}$ was found for 
higher mobility samples).  This is consistent with former experiments, 
where a $g$-factor enhancement at odd $\nu_{0}$ by about one order of 
magnitude (compared to the GaAs bare value $g_{0} = 0.44$) was 
reported.\cite{gexperiments} Very recently, however, one 
group\cite{gconstant} found an enhanced Land\'{e} factor $g^{*} 
\approx 5.2$, but with a spin gap proportional to $B$.  This can not 
be explained by the model of exchange-enhancement.

Furthermore, in agreement with a recent 
prediction\cite{PS-pre} for high mobility samples with long-range 
impurity potential, the prefactors $\sigma_{\mathrm{xx}}^0$ in 
Eq.~(\ref{EqTactiv}) were closely approaching the universal value 
$2e^2/h$ at $\nu_{0}$: the mean value of all 
$\sigma_{\mathrm{xx}}^0(\nu_{0}=2, 3, 4)$ and from all investigated 
samples was $(2.02 \pm 0.11) e^2/h$.  However, at $|\nu - \nu_{0}| 
\gtrsim 0.05$ around even $\nu_{0}$, the prefactors unexpectedly 
dropped by about one order of magnitude.  While an other group 
reported similar behaviour\cite{Svoboda-hop} and attributed it to a 
contribution of VRH conduction, we have doubts about this 
interpretation for such elevated temperatures up to 10 K.  We always 
observed one single exponential slope at intermediate temperatures 
and, what we consider as the hopping contribution, a clear upward 
curvature from the Arrhenius Law at $T \lesssim 1\ \mathrm{K}$ (cf.  
Fig.~1 in Ref.~\onlinecite{ep2ds} and discussion in Sec.~\ref{subVRH} 
of this paper).  Our data at 1 K $< T <$ 10 K and away from $\nu_{0}$ 
are rather consistent with the picture of conduction via extended 
states in the case of a short-range impurity potential,\cite{PS-short} 
although we currently do not understand the reason for the abrupt 
regime crossover.  A possible reason for the reduced prefactors is 
also an effective temperature dependence in the activation energy due 
to the adjustment of $E_{\mathrm{F}}$, in order to keep the number of 
particles constant.  As the Fermi level moves away from the nearest LL 
with increasing temperature, the observed prefactors become smaller 
than in the case of a temperature independent $E_{\mathrm{F}}$.  This 
effect obviously doesn't occur in the special case of electron-hole 
symmetry with $E_{\mathrm{F}}$ in the middle between two LLs (minimum 
of the DOS), yielding the observed universal $\sigma_{\mathrm{xx}}^0$ 
values.

Irrespective of the reasons for the prefactor behaviour, the narrow 
range around $\nu_{0}$, where we have found $\sigma_{\mathrm{xx}}^0 
\simeq 2e^2/h$, can, however, explain the scattering of prefactor values 
experimentally observed by other 
groups.\cite{UniPreFac}

\begin{figure}
\centerline{\psfig{figure=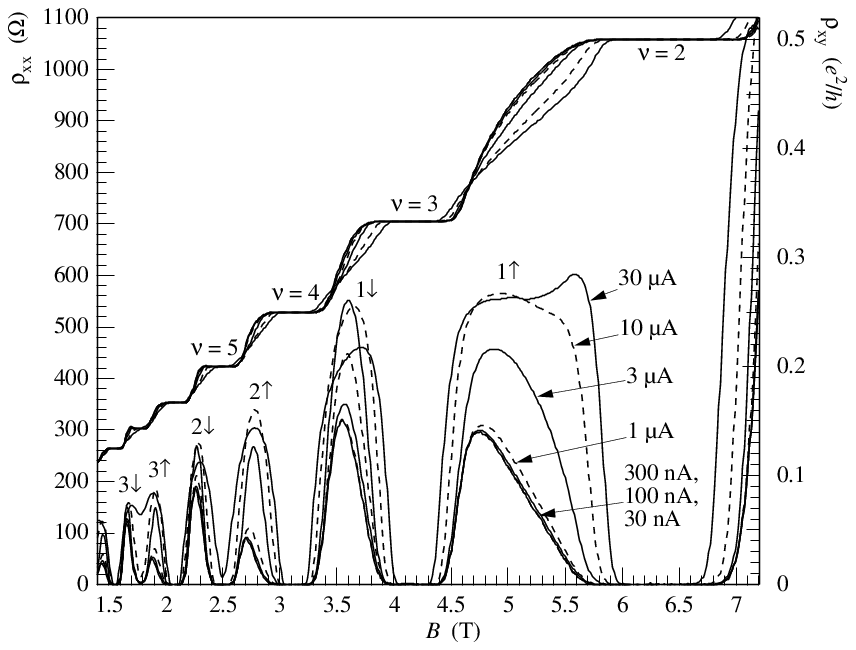}}
\caption{SdH oscillations $\rho_{\mathrm{xx}}(B)$ and Hall 
resistivities $\rho_{\mathrm{xy}}(B)$ at $T \simeq 0.3$~K for 
different bias currents (sample A).  The $\rho_{\mathrm{xx}}$ peaks 
are labeled with their corresponding LL index and the spin 
polarization, the $\rho_{\mathrm{xy}}$ plateaus with the filling 
factor $\nu_{0}$.} \label{figSdH}
\vskip1pc

\centerline{\psfig{figure=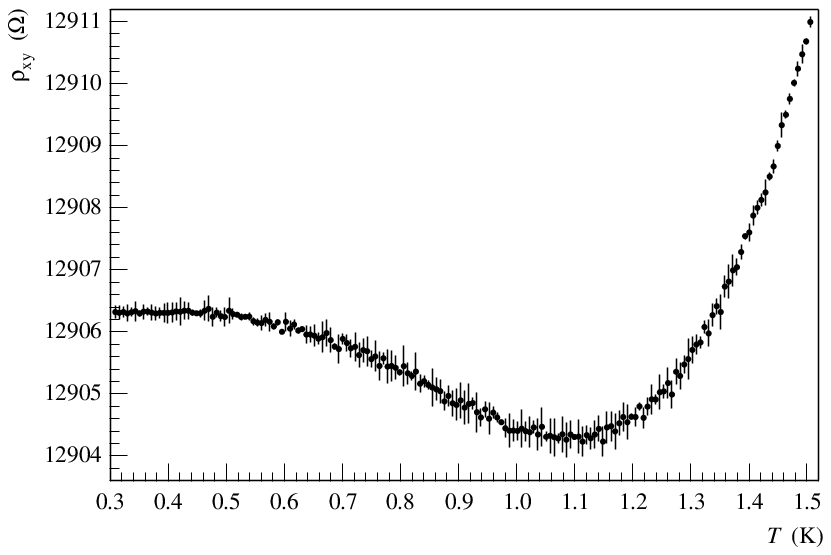}}
\caption{Temperature dependent transverse resistivity 
$\rho_{\mathrm{xy}}(T)$ on the high-$B$ plateau side.  The dip around 
$T \approx 1$~K is due to a geometrical mixing of $V_{\mathrm{x}}$ 
into $V_{\mathrm{H}}$.} \label{figmix}
\vskip1pc

\centerline{\psfig{figure=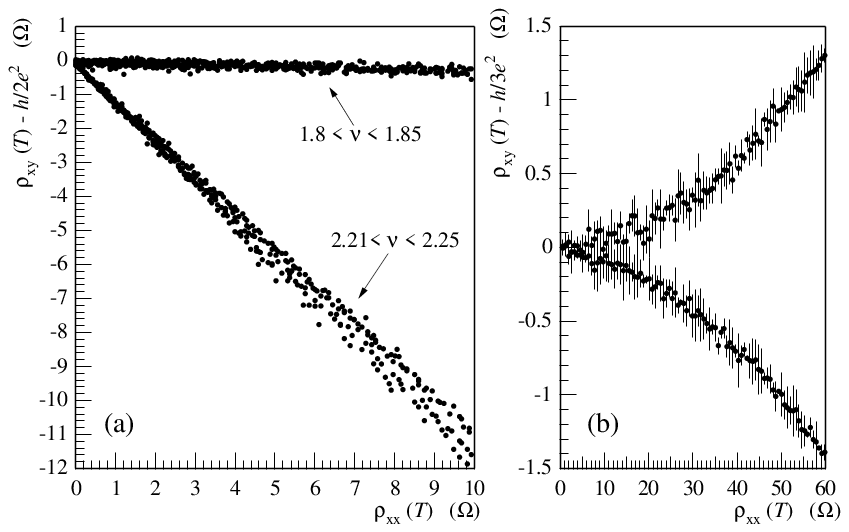}}
\caption{Left (a): temperature-driven flow lines of 
$\delta\rho_{\mathrm{xy}}(T)$ versus $\rho_{\mathrm{xx}}(T)$ at the 
low-$B$ ($\nu \approx 2.23$) and the high-$B$ ($\nu \approx 1.82$) 
plateau side for low $\rho_{\mathrm{xx}}$ values (sample B).  Right 
(b): temperature-driven resistivities as in (a), but for smaller 
$| E_{\mathrm{F}} - E_{\mathrm{c}} |$ (on the third plateau at 
$\nu = 3.06$ and $\nu = 2.94$).  In this case $\rho_{\mathrm{xy}}(T)$ 
is corrected to account for the geometrical mixing effect, according 
to Eq.~(\ref{Eqmix}).} \label{figrxyrxx}
\vskip1pc

\centerline{\psfig{figure=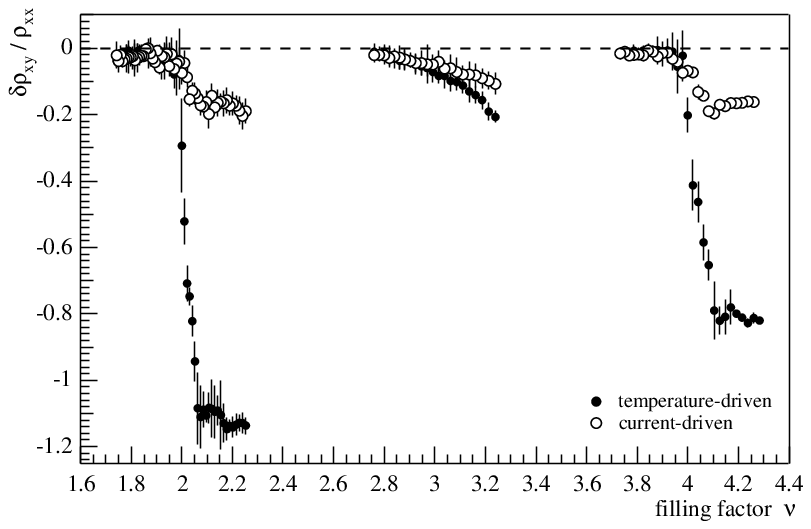}}
\caption{Fit results for the slopes from the linearly related 
$\delta\rho_{\mathrm{xy}} / \rho_{\mathrm{xx}}$ (sample B).  The full 
points {\large $\bullet$} are the temperature-driven data from 
experiment (a), the open circles {\large $\circ$} are from the current 
dependent experiment (b).} \label{figslopes}
\vskip1pc

\centerline{\psfig{figure=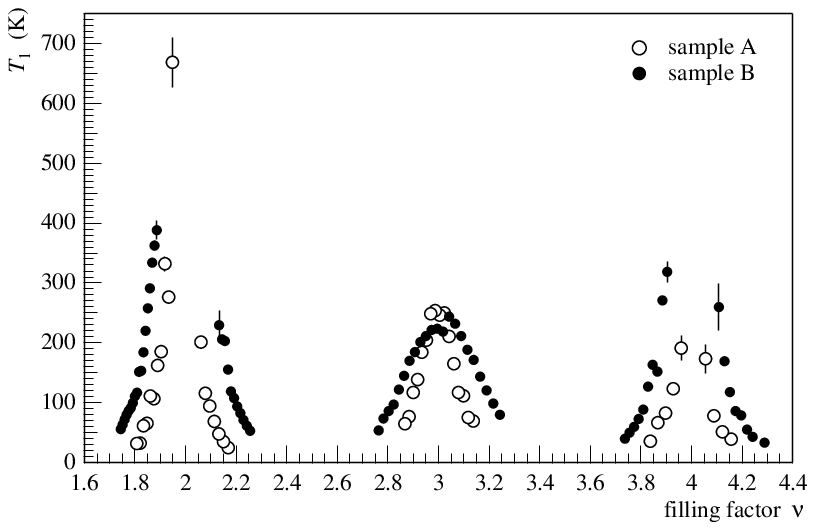}}
\caption{Characteristic hopping temperature $T_{1}$, according to 
Eqs.~(\ref{EqVRH},\ref{EqT1}), and extracted from the experimental data 
in the VRH regime at $T \protect\lesssim 1$ K.  The best fit to the 
data was found with a prefactor proportional to $1/T$.  Around even 
$\nu_{0}$, no data points are available due to unmeasurably small 
resistivities $\rho_{\mathrm{xx}}$ at low $T$.} \label{T1plot} 
\vskip1pc

\centerline{\psfig{figure=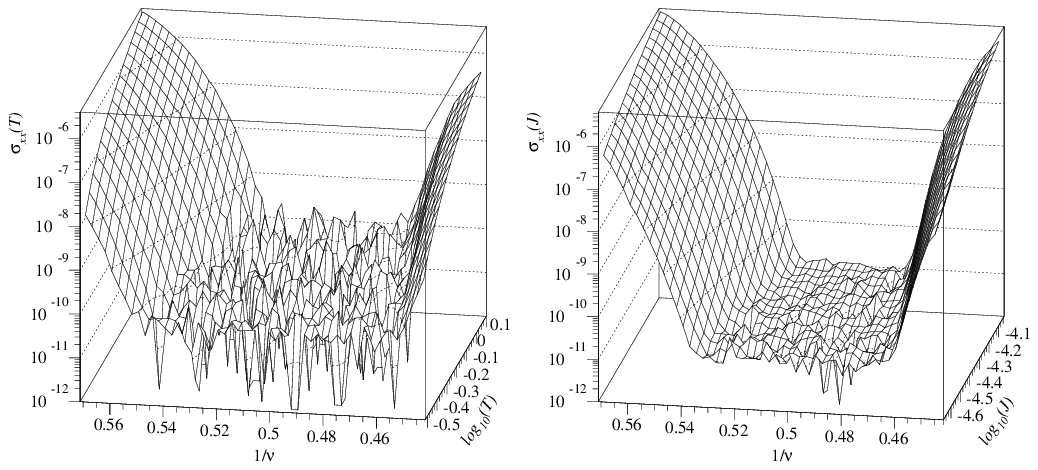}}
\caption{Surface plots of the measured $\sigma_{\mathrm{xx}}$ on the 
second plateau as a function of temperature $T$ (left) or current $J$ 
(right).  The ranges were chosen to show comparable 
$\sigma_{\mathrm{xx}}$ and to correspond to the VRH regime.  The 
saturation of $\sigma_{\mathrm{xx}}$ around the plateau centre 
corresponds to experimental noise, which is decreasing in the 
$\sigma_{\mathrm{xx}}(J)$ plot due to higher currents.} \label{fig3D}
\vskip1pc

\centerline{\psfig{figure=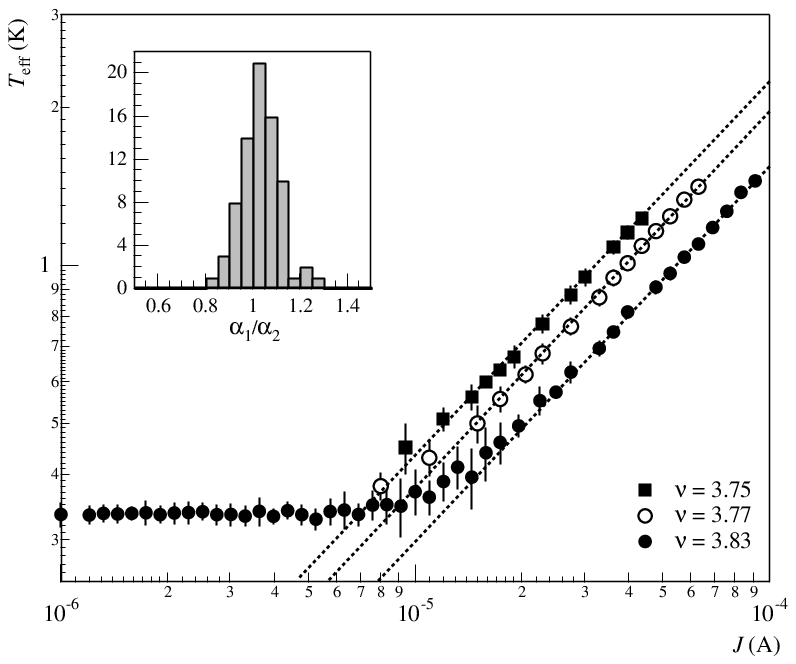}}
\caption{Effective electron temperature $T_{\mathrm{eff}}$ versus 
current $J$ in log-log representation, for three different filling 
factors.  The data at low $J$, where $T_{\mathrm{eff}}$ saturates at 
about 0.32 K (which is the $^3$He bath temperature), correspond to the 
range where $\sigma_{\mathrm{xx}}$ is in the experimental noise.  The 
dotted lines are linear fits to the points at higher $J$.  The inset 
shows the histogram of all slope values, with $\alpha_{1}/\alpha_{2} = 
1.005 \pm 0.096$.} \label{figTeff}
\vskip1pc

\centerline{\psfig{figure=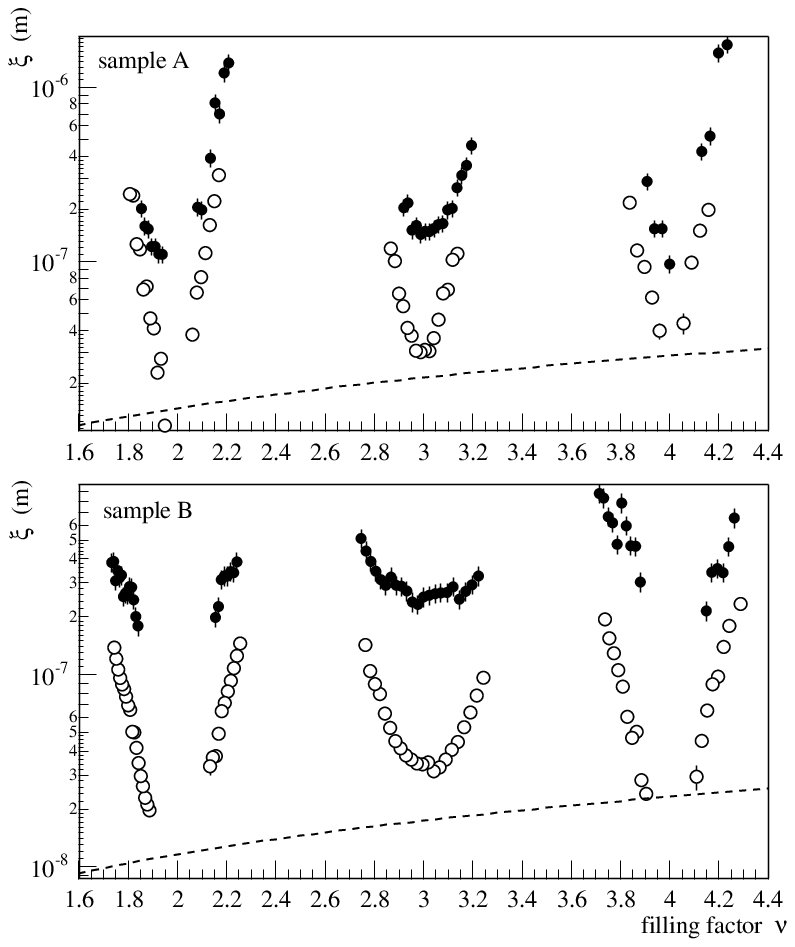}}
\caption{Localization length $\xi$ as a function of $\nu$ obtained from 
the two methods as discussed in the text.  The open circles {\large 
$\circ$} are calculated from the characteristic hopping temperature 
$T_{1}$ (Fig.~\ref{T1plot}), and with Eq.~(\ref{EqT1}).  The full dots 
{\large $\bullet$} show the results from the effective electron 
temperature analysis (Sec.~\ref{subNonohmic}) of non-Ohmic transport 
data.  The dashed lines correspond to the classical cyclotron radius 
$R_{\mathrm{c}}$.  Upper and lower figure are for sample A and B, 
respectively.} \label{figxi}
\vskip1pc

\centerline{\psfig{figure=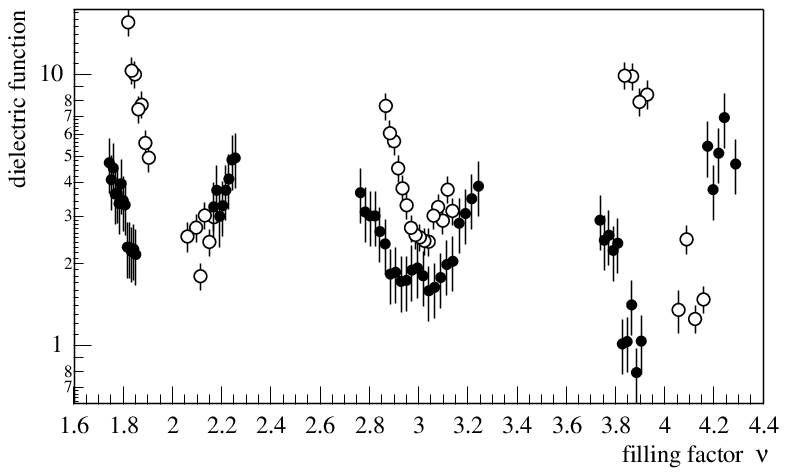}}
\caption{Dielectric function, given by the ratio 
$\epsilon_{\mathrm{r}} \propto \xi(T_{1}) / \xi(T_{\mathrm{eff}})$, 
versus filling factor $\nu$.  Open circles {\large $\circ$} and full 
dots {\large $\bullet$} represent the results for sample A and B, 
respectively.  The divergence of the ratio with decreasing distance 
between Fermi level and the LL centre implicitly suggests a divergence 
of the dielectric function $\epsilon_{\mathrm{r}}(\nu)$.}
\label{epsilondiv}
\vskip1pc

\end{figure}

The investigation of the thermally activated behaviour of the 
transverse resistivity $\rho_{\mathrm{xy}}(T)$ (or conductivity) is 
complicated by a mixing of longitudinal voltage $V_{\mathrm{x}}$ into the Hall 
voltage $V_{\mathrm{H}}$.  Figure \ref{figmix} shows a measurement of 
$\rho_{\mathrm{xy}}(T)$ on the high-$B$ plateau side ($\nu \approx 
1.8$).  The observed decrease of $\rho_{\mathrm{xy}}(T)$ is 
independent of current polarity, magnetic field direction and 
contacts (for the same sample).  It is explained by a geometrical 
mixing\cite{mixing} of $V_{\mathrm{x}}$ over the finite probe arm 
width $w_{\mathrm{p}}$ into $V_{\mathrm{H}}$, yielding an effectively 
measured
\begin{equation}
  \rho_{\mathrm{xy}}^{\mathrm{meas}}(T) = \rho_{\mathrm{xy}}(T) - 
  \frac{w_{\mathrm{p}}}{L_{\mathrm{y}}} \rho_{\mathrm{xx}}(T).
  \label{Eqmix}
\end{equation}
The geometrical ratio is $w_{\mathrm{p}}/L_{\mathrm{y}} = 3 \ldots 
14\%$ for our different samples, and Eq.~(\ref{Eqmix}) satisfactorily
accounts for the observed mixing effect within experimental 
uncertainties.  However, a more general and quantitatively more 
accurate measure of the $\rho_{\mathrm{xy}}(T)$ behaviour can be 
acquired by plotting $\rho_{\mathrm{xy}}(T) = h/e^2 \nu_{0} + 
\delta\rho_{\mathrm{xy}}(T)$ versus $\rho_{\mathrm{xx}}(T)$.  In case 
the Fermi energy $E_{\mathrm{F}}$ is far enough from the percolation 
level $E_{\mathrm{c}}$ and $T$ is not too high (but still above the 
VRH regime), i.e.  low $\rho_{\mathrm{xx}}$ values, we observe 
strictly linear behaviour of the temperature-driven 
$\delta\rho_{\mathrm{xy}}/ \rho_{\mathrm{xx}}\, (\sim - 
\delta\sigma_{\mathrm{xy}}/ \sigma_{\mathrm{xx}})$ typically over 
three decades in $\rho_{\mathrm{xx}}$, as shown in 
Fig.~\ref{figrxyrxx}(a).  While more complicated temperature 
dependencies of the prefactors in the thermal activation formulae were 
predicted,\cite{FPS} our results show that the temperature-driven 
dependence $\delta\rho_{\mathrm{xy}}(\rho_{\mathrm{xx}})$ is dominated 
by the exponential term $\exp(-T_{0}/T)$ in the considered regime.  
For the case of higher temperatures and/or decreasing $T_{0}$, 
crossover to a quadratic dependence $\delta\rho_{\mathrm{xy}} \propto 
\rho_{\mathrm{xx}}^2$ was observed, in agreement with finite-size 
scaling theories\cite{Huckestein} and the so-called ``semicircle 
rule''.\cite{semicircle} A measurement of activated behaviour in this 
regime, taken on the third plateau, is shown in 
Fig.~\ref{figrxyrxx}(b).  In this case, $E_{\mathrm{F}}$ is in the spin 
gap and $T_{0}$ is consequently lower.  Also, $\rho_{\mathrm{xy}}$ 
is corrected according to Eq.~(\ref{Eqmix}) to account for the mixing 
effect.

Now we want to draw more attention to the former case of linearly 
related $\delta\rho_{\mathrm{xy}}(\rho_{\mathrm{xx}})$.  The slopes of 
the temperature-driven $\delta\rho_{\mathrm{xy}}(\rho_{\mathrm{xx}})$, 
obtained from linear fits to the measured data points, are shown in 
Fig.~\ref{figslopes} as a function of $\nu$ (full points).  The 
results on the even numbered plateaus show a strong asymmetry with 
respect to the plateau centre [cf.~also Fig.~\ref{figrxyrxx}(a)].  
This observation was reproduced with all our investigated high 
mobility GaAs/AlGaAs heterostructures and was independent of current 
polarity, magnetic field direction or contact pairs.  The values on 
the high-$B$ plateau side (low-$\nu$) are close to the correction 
$-(w_{\mathrm{p}}/L_{\mathrm{y}})$ for geometrical mixing 
(\ref{Eqmix}) within experimental uncertainties, as discussed above.  
This result implies that in this plateau range $\rho_{\mathrm{xx}}(T)$ 
is strongly enhanced relative to $\delta\rho_{\mathrm{xy}}(T)$.  On 
the low-$B$ plateau side (high-$\nu$), however, this is not the case: 
slopes of order unity were found.  We attribute this asymmetrical 
behaviour to different longitudinal transport regimes, depending on 
the position of the Fermi energy $E_{\mathrm{F}}$.  This leads to a 
picture of either dominantly percolating empty or percolating full 
transport.\cite{Cooper} In the percolating empty regime 
($E_{\mathrm{F}}$ in the low energy LL tail), scattering occurs only 
between edge channels.  In contrast, the interplay between edge and 
bulk states in the percolating full regime ($E_{\mathrm{F}}$ in the 
high energy LL tail) leads to an increased backscattering probability 
with enhanced resistivities measured.  Observation of similar 
asymmetrical behaviour and a model of distinct transport regimes have 
recently been reported.\cite{TwoReg} The interpretation of different 
longitudinal transport regimes is also consistent with the 
experimentally evident asymmetries of the thermal activation 
prefactors\cite{ep2ds} on even numbered plateaus, where \[ 
\frac{\sigma_{\mathrm{xx}}^0(\nu_{0}-\delta\nu)} 
{\sigma_{\mathrm{xx}}^0(\nu_{0}+\delta\nu)} \approx 6 \] was observed 
[see Fig.~3(b) in Ref.~\onlinecite{ep2ds}].  Besides, this may also 
account for the peculiar discontinuity of the prefactor 
$\sigma_{\mathrm{xx}}^0$ at $\nu = 3$, as we observed in some of our 
samples.  We want to stress again that the transport phenomena 
discussed here are not to be confused with experimental observations 
of the anomalous QHE with nonideal contacts, which selectively probe 
only some poorly equilibrated edge channels.  Our data neither depends 
on the longitudinal probe distance, nor are the results compatible 
with a temperature dependent equilibration length.\cite{equilength} 
The extrapolated Hall resistivities 
$\rho_{\mathrm{xy}}(\rho_{\mathrm{xx}} \rightarrow 0)$ perfectly 
coincide with the quantized values $h/e^2 \nu_{0}$.

A basic result of the observed proportionality 
$\delta\rho_{\mathrm{xy}}(T) \propto \rho_{\mathrm{xx}}(T)$ is, besides 
the geometrical mixing effect (\ref{Eqmix}), the indirect 
determination of the activation energy in 
$\delta\rho_{\mathrm{xy}}(T)$: both resistivity tensor components 
essentially follow the same exponential behaviour, which implies the 
same activation energy $k_{\mathrm{B}}T_{0}$.  Although this is not a 
spectacular result but rather an experimental confirmation of general 
theoretical consensus, not much significant and conclusive data about the 
activation energy in $\delta\rho_{\mathrm{xy}}(T)$ has been published 
yet.  The difference between $\delta\rho_{\mathrm{xy}}$ and 
$\rho_{\mathrm{xx}}$ for $E_{\mathrm{F}}$ far from $E_{\mathrm{c}}$ 
are concluded to be a consequence of different prefactors only.  The 
prefactors themselves depend on $\nu$ (Fig.~\ref{figslopes}).

At this point we should emphasize again that the temperature-driven 
resistivities shown in Fig.~\ref{figrxyrxx} and the results in 
Fig.~\ref{figslopes} correspond to the temperature range $T \gtrsim 1$ 
K where transport is dominated by conduction via electrons, which are 
thermally activated to extended states.  At lower temperatures, where 
VRH becomes important, the temperature-driven slopes 
$\delta\rho_{\mathrm{xy}} / \rho_{\mathrm{xx}}$ were found to be equal 
to the ratio $-(w_{\mathrm{p}}/L_{\mathrm{y}})$ for almost our entire 
plateau range, i.e.  only due to the geometrical mixing effect.  This 
is consistent with the theoretical prediction\cite{Wysokinski} of a 
negligibly small VRH contribution to $\delta\rho_{\mathrm{xy}}$ 
compared to $\rho_{\mathrm{xx}}$.

Results from current-driven resistivities obtained in experiment (b), 
i.e.  in the non-Ohmic regime, are also shown in Fig.~\ref{figslopes} 
and will be discussed in Sec.~\ref{subNonohmic}.

\subsection{Variable-range hopping}
\label{subVRH}
As mentioned in the previous section, the temperature dependent 
longitudinal conductivities were observed to deviate from the simple 
exponential Arrhenius behaviour (\ref{EqTactiv}) at low temperatures 
$T \lesssim 1$ K.  In this range, where a contribution of VRH 
conduction according to Eq.~(\ref{Eqgenhop}) is considered, the best 
fit to our experimental data was obtained with $\alpha=\frac{1}{2}$ 
and a temperature dependent prefactor 
$\sigma_{\mathrm{xx}}^{\mathrm{T}} \propto 1/T$, in agreement with 
other experiments.\cite{Eberthopping,Briggs,KochHop} The 
extracted characteristic hopping temperature $T_{1}$ is shown in 
Fig.~\ref{T1plot} as a function of $\nu$.  However, due to our rather 
limited low-$T$ range, our fitting procedure is not very sensitive to the 
prefactor behaviour, and we can not conclusively rule out other 
temperature dependencies of the prefactors.  In the case of fitting with a 
temperature independent $\sigma_{\mathrm{xx}}^{\mathrm{T}}$ (and 
accepting a slightly worse agreement between the fit function and the data 
points), we obtain values for $T_{1}$ which are typically $20\%$
larger than those shown in Fig.~\ref{T1plot}.  Our $T_{1}$ results can 
therefore be considered to have a maximum uncertainty of this 
magnitude.

According to the theory of Ref.~\onlinecite{Ono}, the preexponential 
factor $\sigma_{\mathrm{xx}}^{\mathrm{T}}  = e^2 \gamma_{0} / 
k_{\mathrm{B}}T$ contains the material 
parameter $\gamma_{0}$, which is essentially a material constant, 
depending on the electron-phonon coupling strength.  We observed, 
however, a pronounced asymmetry of $\gamma_{0}$ with respect to the 
plateau centres (not shown here), similar to the reported prefactor 
asymmetries of thermally activated $\sigma_{\mathrm{xx}}$ at higher 
$T$.  On the low-$\nu$ plateau side, the values were about one order 
of magnitude larger than on the high-$\nu$ side.  This result can 
certainly not be understood in terms of electron-phonon scattering 
only.  We have estimated the density of states (DOS) of the 
two-dimensional electron gas either by means of the activation energies 
$k_{\mathrm{B}}T_{0}(\nu)$ and with the procedure proposed in 
Ref.~\onlinecite{Weiss}, or from the characteristic temperatures $T_{1}$ 
according to the hopping model in Ref.~\onlinecite{Ono}.  As mentioned 
before, the unequal contribution of attractive and repulsive 
scatterers in high mobility heterostructures may be taken to be 
responsible for the asymmetries of the DOS,\cite{HaugAsym} having an 
influence e.g.  on the shape of SdH peaks.  Within our $\nu$ range, 
the DOS does not show any significant asymmetries that could account 
for the dramatic prefactor behaviour.  We interpret the systematic 
asymmetries of the prefactors again with different transport regimes, 
depending on the position of $E_{\mathrm{F}}$ in the LL tail relative 
to the mobility edge (as discussed in Sec.~\ref{subTactiv}).

With respect to our studies, there is only one published 
experiment\cite{Eberthopping} up to now with a thorough quantitative 
investigation and useful data on hopping transport at the resistivity 
minima.  From their analysis based on the hopping theory of 
Ref.~\onlinecite{Ono}, they claim to extract values for $T_{1}$, which are 
more than one order of magnitude too small (implying a much too large 
DOS compared to the zero field value).  In contrast to their results, 
our $T_{1}$ values far in the LL tails are well consistent with the 
mentioned hopping theory.\cite{Ono} However, at the resistivity 
minima, the predictions for $T_{1}$ in Ref.~\onlinecite{Ono} virtually 
coincide with those in Refs.~\onlinecite{ElProp,Efros} [i.e.  
Eq.~(\ref{EqT1})] for our samples at low $\nu$.  Hence, from the 
results around the plateau centres, we are not able to give precedence 
to either of the two models.  On the other hand, however, the theory 
with the assumption\cite{Ono} of Gaussian localization of the electron 
wavefunction $\psi(r) \propto 
\mathrm{e}^{-r^{2}/4\ell_{\mathrm{B}}^{2}}$ has been 
criticized\cite{PS-peak} with the argument that the tails of the 
wavefunction have a simple exponential form\cite{ExpWfun} $\psi(r) 
\propto \mathrm{e}^{-|r|/\xi}$.  Furthermore, it was mentioned before 
that the observed prefactor asymmetries are not consistent with the 
theoretical prediction in Ref.~\onlinecite{Ono}.  Therefore, we will 
base our further analysis on the VRH theory as developed in 
Refs.~\onlinecite{ElProp,Efros,PS-peak}, together with 
Eq.~(\ref{EqT1}).

\subsection{Non-Ohmic transport and effective electron temperature}
\label{subNonohmic}
Several studies on the current dependent non-Ohmic transport were 
previously published, as mentioned in the introduction.  Before examining the 
applicability of the models to our experimental results, we shall 
first consider the current-driven relation between the measured 
$\rho_{\mathrm{xy}}(J)$ and $\rho_{\mathrm{xx}}(J)$, obtained from 
experiment (b).  Similarly to the temperature-driven resistivities 
(Sec.~\ref{subTactiv}), we found essentially linear behaviour of the 
current-driven $\rho_{\mathrm{xy}}(\rho_{\mathrm{xx}})$ typically over 
three decades of $\rho_{\mathrm{xx}}$.  The results for the slopes are 
shown in Fig.~\ref{figslopes} (open circles).  The same asymmetries 
were observed as in the case of experiment (a), but with significantly 
smaller values.  On the low-$\nu$ plateau side, the slope values from 
both experiments coincide, i.e.  they correspond to the geometrical 
mixing coefficient $-(w_{\mathrm{p}}/L_{\mathrm{y}})$, according to 
Eq.~(\ref{Eqmix}).  Thus, we can conclude nothing about a possible 
difference in the actual behaviour and the relations between the 
resistivities in that range.  On the high-$\nu$ side, however, the 
slope values typically differ by a factor of 4 to 6.  The observed 
linear behaviour of $\rho_{\mathrm{xy}}(\rho_{\mathrm{xx}})$ in both 
experiments (a) and (b) implies that the resistivities are dominated 
by the exponential term in Eqs.~(\ref{EqTactiv}) and (\ref{EqJhop}) 
for the dependence on $T$ or $J$, respectively.  Therefore, it must be 
the prefactors which differ in cases (a) and (b).  One should keep in 
mind that the temperature-driven slopes represent thermally activated 
transport, whereas the current-driven results are obtained in the VRH 
regime.  Therefore, we interpret the discrepancy in the slope values 
from the two experiments as a difference in the prefactors of the 
transverse resistivities due to different transport processes.  
Moreover, it was mentioned in Sec.~\ref{subTactiv} that in the low-$T$ 
range of VRH conduction, the deviations of 
$\delta\rho_{\mathrm{xy}}(T)$ were much smaller than 
$\rho_{\mathrm{xx}}(T)$, supporting the idea of transport regime 
dependent $\delta\rho_{\mathrm{xy}}$ prefactors.

It is interesting to directly compare the conductivities obtained from both 
experiments (a) and (b).  The $\sigma_{\mathrm{xx}}$ values measured 
on the second plateau as a function of temperature and of current are 
shown in Fig.~\ref{fig3D}.  The ranges of $T$ and $J$ were chosen to 
show comparable $\sigma_{\mathrm{xx}}$ values.  Also, the $T$-range 
corresponds in this case to low temperatures with VRH conduction.  The 
two plots are qualitatively very similar.  We can now relate the 
quantities from both experiments by comparing the measured 
$\sigma_{\mathrm{xx}}(T) \equiv \sigma_{\mathrm{xx}}(J)$ point by 
point.  This yields an effective electron temperature 
$T_{\mathrm{eff}} = T\bigl(\sigma_{\mathrm{xx}}(J)\bigr)$ for a given 
current $J$, as discussed in Sec.~\ref{foreplay}.  Results from this 
analysis for $T_{\mathrm{eff}}(J)$ at three different $\nu$ are shown 
in Fig.~\ref{figTeff}.  At low current levels, the measured 
conductivities are below the experimental noise, resulting in an 
artificially saturated $T_{\mathrm{eff}} \approx 320$ mK, which is 
simply the $^3$He bath temperature.  At higher $J$, however, a clear 
$T_{\mathrm{eff}}(J)$ dependence is observed.  Data analysis was 
performed in this way for all measured $\nu$.  The extracted effective 
temperatures were found to never exceed the range $T_{\mathrm{eff}} = 
0.3 \ldots 1.4$ K.  This justifies a treatment on the basis of hopping 
theory.  If the temperature and current dependent conductivities obey 
the laws (\ref{EqVRH}) and (\ref{EqJhop}), respectively, we are able 
to relate the measurements according to
\begin{equation}
  \left(\frac{{\mathcal{E}}_{1}}{{\mathcal{E}}_{\mathrm{H}}}\right)^{\alpha_{1}} = 
  \left(\frac{T_{1}}{T}\right)^{\alpha_{2}} + 
  \ln\frac{\sigma_{\mathrm{xx}}^{\mathrm{J}}} 
  {\sigma_{\mathrm{xx}}^{\mathrm{T}}},
  \label{Eqrelate}
\end{equation}
with $\alpha_{1},\alpha_{2} = \frac{1}{2}$ expected.  To test this 
condition we applied a linear fit to the points at high $J$ in the 
log-log plot (dotted lines in Fig.~\ref{figTeff}).  The values of the 
slopes of the straight lines for all considered $\nu$ and for both 
samples are shown in the histogram (inset in Fig.~\ref{figTeff}).  
They correspond to the ratio $\alpha_{1}/\alpha_{2}$ and were found to 
be $\alpha_{1}/\alpha_{2}=1.005 \pm 0.096$.  While $\alpha_{2} = 
\frac{1}{2}$ has already been confirmed in Sec.~\ref{subVRH}, this 
result supports the interpretation of transport based on 
Eqs.~(\ref{EqVRH}) and (\ref{EqJhop}).  Finally, the position of the 
fitted straight lines yields the results for 
$T_{1}(\nu)/{\mathcal{E}}_{1}(\nu)$, which gives information on the 
localization length $\xi(\nu)$ [cf.  Eq.~(\ref{EHchar})].  This is 
considered in Sec.~\ref{discuss}.

In the above argument we have neglected the additional term
$\ln(\sigma_{\mathrm{xx}}^{\mathrm{J}} / \sigma_{\mathrm{xx}}^{\mathrm{T}})$ 
in Eq.~(\ref{Eqrelate}).  It can, however, easily be seen from 
Fig.~\ref{T1plot} that in most cases $\sqrt{T_{1}/T}$ is much larger than 
$\ln(\sigma_{\mathrm{xx}}^{\mathrm{J}} / \sigma_{\mathrm{xx}}^{\mathrm{T}})$ 
even for possible differences of the prefactors by one order of 
magnitude (cf.  discussion above).  If that would not 
be true, no straight lines in $\log(T_{\mathrm{eff}})$ versus $\log(J)$ 
should have been observed, and slopes different from one should 
result.  To be complete, this was indeed the case at certain $\nu$ values 
far away from the plateau centre, where $T_{1}$ was very low.  Such 
data was excluded from further analysis.

The same argument of negligible $\ln(\sigma_{\mathrm{xx}}^{\mathrm{J}} 
/ \sigma_{\mathrm{xx}}^{\mathrm{T}})$ also applies to the case where 
the behaviour of the prefactors is not exactly known.  Therefore, and 
this is most important, the experimental method and the way of data 
analysis discussed in this section allow to investigate some fundamental 
relations in quantum transport even if certain parameters (like the 
functional behaviour of the prefactors) are not perfectly determined.

\section{Analysis and discussion}
\label{discuss}
In this section we will mainly focus on the results from the 
temperature and current dependent VRH conductivity measurements 
presented in Secs.~\ref{subVRH} and \ref{subNonohmic}.  Both 
experimental methods allow to extract the localization length $\xi$.  
The values for $\xi(T_{1})$ and $\xi\bigl(T_{\mathrm{eff}}(J)\bigr)$, 
calculated with Eqs.~(\ref{EqT1}) and (\ref{EHchar}), respectively, 
are shown in Fig.~\ref{figxi}.  A constant dielectric function 
$\epsilon_{\mathrm{r}} = 13$ was assumed in the case of $\xi(T_{1})$.  
Close to the even numbered plateau centres no data points are available 
because of unmeasurably small resistivities.  The divergence of 
$\xi(\nu)$ for $\nu$ approaching half filling fractions is well 
understood within the model of two-dimensional electron localization.  
Close to the LL centre, the behaviour of $\xi(\nu)$ is expected to 
follow the power law (\ref{xidiverg}).  For large energy separation 
$|E_{\mathrm{c}} - E_{\mathrm{F}}|$, i.e.  $E_{\mathrm{F}}$ deep in 
the mobility gap between two LLs, the lack of knowledge about the 
exact form of the density of states doesn't allow to explicitly deduce 
the functional behaviour of the measured $\xi$.  However, it is 
expected\cite{loclength} that $\xi$ approaches a length close to the 
classical cyclotron radius $R_{\mathrm{c}} = \nu / \sqrt{2\pi 
n_{\mathrm{e}}}$ for $\nu \rightarrow \nu_{0}$.  This is indeed the 
case for $\xi(T_{1})$.  Compared to this prediction, the values of 
$\xi\bigl(T_{\mathrm{eff}}(J)\bigr)$ are anomalously large.  The main 
formal difference between the two methods is that Eq.~(\ref{EqT1}) 
includes the dielectric function $\epsilon_{\mathrm{r}}$ (and the 
constant $C$), whereas Eq.~(\ref{EHchar}) is based on the assumption 
of a homogeneous electric field.  Different reasons may be found for 
the discrepancy between the localization lengths obtained from the two 
methods.  The first criticism addresses the assumption of uniform 
electric field.  Several theoretical\cite{DistrTheo} 
and experimental\cite{qRedistr} investigations strongly suggest a 
significant charge and potential redistribution in the quantized 
regime.  While the electrostatic potential in the metallic phase is 
essentially linear, the potential drops for filling factors close to 
integer mainly occur close to the sample edges, leading to strongly 
enhanced local field gradients.  The width of these potential drops 
was observed to depend on external conditions like contacting geometry 
or an additional gate potential.  In spite of theoretically 
predicted\cite{Chklovskii} narrow edge widths of less than 1 $\mu$m, 
there is also experimental evidence of a very wide region of up to 100 
$\mu$m where the dominant potential drop occurs\cite{PotDrop}.  
Concerning the electric field dependent hopping model (\ref{EHchar}), 
this picture leads to an effectively reduced sample width, being 
typically one order of magnitude smaller than the physical widths of 
our Hall bars.  Taking such an estimate into account, the localization 
lengths $\xi\bigl(T_{\mathrm{eff}}(J)\bigr)$ should be reduced by the 
same amount towards $R_{\mathrm{c}}$ and closely approach the values 
of $\xi(T_1)$ at the plateau centres.

Another reason for locally increased electric fields may be found in 
the macroscopically inhomogeneous nature of the 
samples.\cite{inhomohop}  Some 
spatially rare critical key resistances, composing the network of the 
infinite cluster where the current flows in the VRH regime, determine 
in first order the macroscopic resistance of the medium.  While most 
of the potential drop occurs across such key resistances, the local 
field is enhanced there by the ratio of the characteristic distance of 
the critical sites and the hopping length. The latter is 
usually much smaller than the correlation radius of the infinite 
cluster.  Although this effect may alter the effectively 
applied local electric fields, it is not clear how to quantify the 
model in our case.  We realize, however, that both pictures of field 
inhomogeneity induced either at the sample edges or somewhere along 
the current path due to macroscopic impurities tend to correctly 
account for our experimental results.  Our arguments are the 
following: although sample A has the higher electron mobility 
$\mu_{\mathrm{e}}$, it shows a significantly lower critical breakdown 
current $J_{\mathrm{c}}$ by about one order of magnitude compared to 
sample B (the latter having $J_{\mathrm{c}} \simeq$ 600 $\mu$A at $\nu 
= 2.0$).  Hence, in spite of sample A's ``higher electronic quality'', 
it is less robust against increased ${\mathcal{E}}_{\mathrm{H}}$.  
This is a consequence of a higher degree of macroscopic density 
fluctuation or large scale impurities in sample A, governing the transport 
properties under critical conditions, \emph{in addition} to the
smooth long-range random potential.  The picture of the breakdown 
mechanism with large scale distributed random impurities was recently 
investigated in Ref.~\onlinecite{Breakdown}.  Those 
experiments confirmed the idea of locally triggered breakdown at rare 
critical sites with enhanced electric field.  The larger the number of 
macroscopic impurities (or the larger the sample), the higher the 
probability to exceed the critical threshold at such a site.  Taking 
this into account in the context of the discussed locally increased 
field within the VRH model, one would expect better agreement between 
${\mathcal{E}}_{\mathrm{H}}$ and the average effective local field for 
sample A.  This is indeed consistent with our 
$\xi\bigl(T_{\mathrm{eff}}(J)\bigr)$ results, which show lower values 
for sample A, being in better agreement with predictions and the 
temperature dependent VRH experiment.  On the other hand, sample A is 
half as wide as sample B, i.e.  the physical width is closer to the 
potential drop width in the scenario of increased field at the sample edges.  
This again leads to lower $\xi\bigl(T_{\mathrm{eff}}(J)\bigr)$ for 
sample A, as observed in the data.  To conclude this discussion, more 
experiments are needed to distinguish the field enhancement mechanisms 
and to answer the question about the dominant contribution.

Next, we want to comment on the $\xi(T_{1})$ values related to the 
temperature dependent experiment.  Although the dielectric function 
$\epsilon_{\mathrm{r}}$ appearing in Eq.~(\ref{EqT1}) is usually assumed to 
be constant in the two-dimensional case, this is a crude approximation 
and is neither theoretically nor experimentally established.  In real 
systems, the dielectric function is believed to diverge in the 
three-dimensional case when approaching the LL centre.\cite{xidiv} 
Drawing the analogy for the quantum Hall effect in two dimensions, the 
dielectric function grows like $\epsilon_{\mathrm{r}} \propto 
\xi^{\beta}$ with $0 \leq \beta \leq 1$.\cite{PS-peak} Since our 
experiment in the VRH temperature regime actually measures 
$\epsilon_{\mathrm{r}}\,\xi(T_{1})$ as a function of $\nu$ according 
to Eq.~(\ref{EqT1}), the ratio between our $\xi(T_{1})$ (where a 
constant $\epsilon_{\mathrm{r}}=13$ was assumed before) and 
$\xi\bigl(T_{\mathrm{eff}}(J)\bigr)$ gives the relative behaviour of 
$\epsilon_{\mathrm{r}}(\nu)$, under the assumption that everything 
else varies at a negligible level relative to $\xi(\nu)$.  As far as 
concerning the local electric field distribution or the field 
enhancement mechanisms, the latter condition of invariance is not 
necessarily true.  We will, however, shortly give an intuitive 
argument justifying our approach.  The dielectric function
\begin{equation}
  \epsilon_{\mathrm{r}}(\nu) = \frac{C e^3 {\mathcal{E}}_{1}}{8\pi \epsilon_{0} 
  k_{\mathrm{B}}^2 T_{1}^2} ,
  \label{Eqepsilon}
\end{equation}
deduced from the experimental data and based on 
Eqs.~(\ref{EqT1},\ref{EHchar}), is plotted in Fig.~\ref{epsilondiv}.  
The reduced $\epsilon_{\mathrm{r}}$ around integer $\nu$ compared to 
the GaAs bare value is due to the underestimate of the electric field 
by about one order of magnitude, as discussed above.  The systematic 
divergence of $\epsilon_{\mathrm{r}}(\nu)$ with decreasing $| 
E_{\mathrm{F}} - E_{\mathrm{c}} |$ is observed here for the first time 
for two-dimensional electron systems in the quantum Hall regime.  For 
sample B, the significant effect appears symmetric with respect to 
integer $\nu_{0}$, whereas the weak asymmetry observed for sample A is 
attributed to a slight shift of the electron densities between the two 
experiments.  A fit of the assumed power law dependence to 
the points $\epsilon_{\mathrm{r}}$ versus 
$\xi\bigl(T_{\mathrm{eff}}(J)\bigr)$ for all $\nu$ yields the exponent 
$\beta = 1.098 \pm 0.096$, independent of LL index.  This result is in 
agreement with theoretical considerations about a filling factor dependent 
dielectric function, although it has not been experimentally observed 
before.

We have argued above that the functional behaviour of the 
experimentally determined $\epsilon_{\mathrm{r}}(\nu)$ reflects the 
divergent dielectric function and not a variation of 
$\xi(T_{\mathrm{eff}})$ due to $\nu$ dependent electric field 
distribution effects.  The reason why we conclude that, is the 
following: the distance $| E_{\mathrm{F}} - E_{\mathrm{c}} |$ in the 
spin gap (third plateau) is small enough (in contrast to plateaus two 
and four) to potentially observe the power law divergence of the 
localization length according to Eq.~(\ref{xidiverg}).  There, the critical 
exponent $\gamma$ was found [from fitting (\ref{xidiverg}) to the 
$\xi(\nu)$ data points] to be $\gamma = 2.29 \pm 0.21$ in the case of 
$\xi\bigl(T_{\mathrm{eff}}(J)\bigr)$ and $\gamma = 4.61 \pm 0.24$ for 
$\xi(T_{1})$.  Hence the critical exponent in the former case is well 
consistent with the theoretically predicted value $\sim 2.3$, whereas 
in the latter case $\gamma$ is too large by a factor of two (we should 
remind, that we are not in the situation here of two strongly overlapping 
spin levels, which might lead to an enhanced critical exponent by a factor 
of two\cite{PS-peak}).  This result justifies the assumption made 
above of negligible effective electric field variation within our 
$\nu$ range.  It rather supports the picture where $\xi(T_{1})$ in 
Fig.~\ref{figxi} actually represents the measurement of 
$\epsilon_{\mathrm{r}}(\nu) \cdot \xi(\nu)$ with 
$\epsilon_{\mathrm{r}} \propto \xi$.  Although this argument is 
consequently based on our experimental results, more measurements on 
extended $\nu$ and temperature ranges are needed to investigate the 
subject (possibly with other high mobility samples), and to confirm 
its implications on the electrical properties in the quantized Hall 
regime.

\section{Conclusions}
\label{conclude}
Results from a large series of transport measurements on quantum Hall 
bars have been reported.  We could clearly distinguish between 
thermally activated transport and such dominated by VRH.  In the 
former case, the longitudinal conductance in our high mobility samples 
well agreed with the Arrhenius Law.  The extrapolated prefactors were 
found to be $(2.02 \pm 0.11) e^{2} /h$ within a narrow range around 
the plateau centres.  Deviations $\delta\rho_{\mathrm{xy}}$ of the 
transverse resistivity from the quantized value were attributed to a 
mixing of $\rho_{\mathrm{xx}}$ into $\rho_{\mathrm{xy}}$ due to finite 
probe arm widths on one hand, and to thermal activation 
$\delta\rho_{\mathrm{xy}} \propto \mathrm{e}^{-T_{0}/T}$ with the 
activation energy $T_{0}$ on the other hand.  The activation energy 
$T_{0}$ was shown to be the same for both $\rho_{\mathrm{xx}}(T)$ and 
$\delta\rho_{\mathrm{xy}}(T)$.  The observed strong asymmetry of 
$\delta\rho_{\mathrm{xy}}/\rho_{\mathrm{xx}}$ with respect to the 
plateau centre was explained with an asymmetry of the 
$\rho_{\mathrm{xx}}$ prefactors due to either percolating full or 
percolating empty transport regimes.

At lower temperatures, both temperature and current dependent 
longitudinal conductivity could be well understood on the basis of a 
VRH theory\cite{ElProp,Efros} taking a Coulomb gap into account.  
Those experiments allowed to determine the localization length 
$\xi(\nu)$ in two different ways.  For the first time, a divergence of 
$\xi$ for $E_{\mathrm{F}}$ approaching the LL centre could be 
demonstrated in the quantum Hall effect over a relatively broad $\nu$ range.  
Inhomogeneous electric field distribution (either due to edge effects 
or macroscopic impurities) was considered to explain discrepancies 
between the two methods.  Most interestingly, our experimental results 
suggest a divergence of the dielectric function $\epsilon_{\mathrm{r}}$.  
First, this divergence was deduced from the ratio $\xi(T_{1}) / 
\xi(T_{\mathrm{eff}})$ of the two differently obtained sets of $\xi$ 
values, according to Eq.~(\ref{Eqepsilon}): $\xi(T_{1})$ contains 
$\epsilon_{\mathrm{r}}$, and the ratio diverges on the third plateau 
like $|\nu - \nu_{\mathrm{c}}|^{-2.3}$ (see Fig.~\ref{epsilondiv}).  
Second, the critical localization exponent was found to be equal to the 
theoretical value $\gamma \simeq 2.3$ for the $\epsilon_{\mathrm{r}}$ 
independent, or twice that value in the case of the 
$\epsilon_{\mathrm{r}}$ dependent model.  Hence, the experiment in the 
latter case measured $\epsilon_{\mathrm{r}} \cdot \xi$ with 
$\epsilon_{\mathrm{r}} \propto \xi$.  We suggest further measurements, 
with the experimental range extended to larger $\nu$ and lower 
temperatures, in order to verify our conclusions and to determine the 
$\xi$ values at the resistivity minima as a function of $\nu$.

\section*{Acknowledgments}
We are grateful to M.  M.  Fogler, B.  Jeanneret, B.  Jeckelmann, L.  
Schweitzer, and B.  I.  Shklovskii for very useful discussions, to U.  
Feller and M.  Ilegems for supporting this work, to H.-J.  B\"uhlmann for 
fabrication of the EPFL samples and to H.  B\"artschi for his 
technical skills.  This work was supported by ISI Foundation and 
ESPRIT `Quantum Hall Effect'.

%

%

\newpage

\onecolumn

\begin{table}
\caption{Electron sheet density $n_{\mathrm{e}}$, mobility 
  $\mu_{\mathrm{e}}$ and spacer thickness $d$ for the two Hall bar 
  GaAs/AlGaAs heterostructures.  The corresponding Fermi wave vector 
  $k_{F}$ and the elastic mean free path $\ell_{\mathrm{mfp}}$ are 
  calculated.  $L_{\mathrm{x,min}}$ is the distance between two 
  consecutive voltage probes, and $L_{\mathrm{y}}$ is the Hall bar 
  width.  Sample A was produced at the Niels Bohr Institute 
  (K$\o$benhavn, Denmark), sample B at EPFL (Lausanne, Switzerland).}

\label{samples}
\begin{tabular}{cccccccc}
      sample & 
      $n_{\mathrm{e}}\ (\mathrm{m}^{-2})$ & 
      $\mu_{\mathrm{e}}\ (\mathrm{T}^{-1})$ & 
      $d\ (\mathrm{nm})$ & 
      $k_{F}\ (\mathrm{m}^{-1})$ &
      $\ell_{\mathrm{mfp}}\ (\mu\mathrm{m})$ & 
      $L_{\mathrm{x,min}}\ (\mathrm{mm})$ & 
      $L_{\mathrm{y}}\ (\mathrm{mm})$ \\
\hline
\textbf{A}: HC$\O$ 130/92  &  $3.09 \times 10^{15}$  &  132  &  25  & 
      $1.39 \times 10^{8}$  &  12.1  &  1.50  &  0.50 \\
\textbf{B}: EPF 277/5      &  $4.74 \times 10^{15}$  &  38.8 &  10  & 
      $1.73 \times 10^{8}$  &   4.4  &  1.25  &  1.00 \\
\end{tabular}
\end{table}

\end{document}